\documentclass{aa}
\usepackage{psfig}
\usepackage{graphicx}
\usepackage{txfonts}
\begin{document}
\title{Quantitative interpretation of the rotation curves of \\ 
   spiral galaxies at redshifts $z \sim 0.7$ and $z \sim 1$} 
   
\author{B. Fuchs\inst{1}\fnmsep\thanks{e-mail:fuchs@ari.uni-heidelberg.de}
          \and
          A. B\"ohm\inst{2}
          \and
          C. M\"ollenhoff\inst{3}
          \and
          B. L. Ziegler\inst{2}}

   \offprints{B. Fuchs}

   \institute{Astronomisches Rechen-Institut,
              M\"onchhofstra{\ss}e 12-14, 69120 Heidelberg, Germany\\
         \and
             Universit\"atssternwarte G\"ottingen,
             Geismarlandstra{\ss}e 11, 37083 G\"ottingen, Germany\\
         \and
             Landessternwarte Heidelberg,
             K\"onigstuhl, 69117 Heidelberg, Germany\\ 
             }

   \date{Received 2004; accepted }

   \abstract{
   We present decompositions of the rotation curves of three spiral galaxies at
   redshifts $z \sim$ 0.7 and 1 into contributions by their bulges, disks, and 
   dark halos, respectively. In order to set constraints on the degeneracy of 
   the decompositions we interpret the morphology of the spiral structures
   quantitatively in the framework of density wave theory. Galaxy models 
   constrained in such a way show that the distant galaxies, which are much
   younger than nearby galaxies, have very likely `maximum disks', i.e. are 
   dominated in their inner parts by baryonic matter. We argue that current 
   theories of the cosmogony of galaxies must allow for these types of galaxies.
   }
   \maketitle
%
\section{Introduction}

The rotation curves of spiral galaxies, in particular the HI -- rotation 
curves, have incontrovertibly revealed that the outer parts of the galaxies are 
dominated by dark matter. For instance, van Albada et al.~(1985) have measured
the rotation curve of NGC\,3198 in HI extending radially outwards to 11 disk
scale lengths, much beyond the optical disk. At such a galactocentric radius the
rotation curve of a single disk would have fallen to half of its peak value at
2.2 radial scale lengths (cf.~equation 5), whereas the observed rotation curve
stays flat. However, there is an intense debate in the literature if and to 
what degree the inner parts of the galaxies, where their optical disks reside, 
are dominated by dark matter as well. The various arguments and the present
state of this so called `maximum disk' problem have been reviewed recently by
Bosma (2004). The reason for this dispute is the notorious degeneracy
of the decomposition of the rotation curves into the contributions by the 
various constituents of the galaxies such as bulge, stellar disk, interstellar
gas disk, and dark halo, respectively. Thus, further constraints on the 
decomposition of the rotation curves are needed. 
Using arguments of the density wave theory of galactic spiral arms Fuchs (2003)
has pointed out that the implied dynamics of the disks might provide such 
constraints for modelling spiral galaxies. Its application to the
nearby bright galaxies NGC\,2985 and NGC\,3198 points towards maximum disk
models in both cases,
i.e.~the contributions of the disks to the rotation curves are at the 
maximum allowed by the data. This would imply that baryons dominate completely
the inner parts of the galaxies. 

If this constraint is applied to the decomposition of the rotation curves
of redshifted galaxies, this allows insight into the cosmogony of spiral
galaxies. For this purpose Fuchs, M\"ollenhoff \& Heidt (1998) have analyzed the
optical
rotation curves of two spiral galaxies at redshifts $z = 0.15$ and $z = 0.48$,
respectively, which had been imaged with HST and showed clearly discernible 
spiral structures. The rotation curves were measured with the Keck telescope
(Vogt et al. 1996). Fuchs et al. (1998) concluded that the disks of these 
galaxies are submaximal. This has to be modified, though, because they assumed 
flat rotation curves for the galaxies. Since the rotation curves span radially 
only two and three radial scale lengths, respectively, the rotation curves are 
still rising over these ranges, which had not been taken into account. The 
disks were modelled as exponential disks, and their rotation 
curves rise until 2.2 radial scale lengths and develop only then a flat
plateau (cf.~equation 5). This affects the density wave theory argument through
the coefficient $X(A)$ in equation (8) below, and, when properly taken into 
account, the disks of the two redshifted galaxies seem to be maximum disks as 
well (Leeuwin \& Fuchs, priv. com.).

%
\begin{figure*}
  \centering
  \vbox{
   \hbox{\hspace{0cm}              
    \psfig{figure=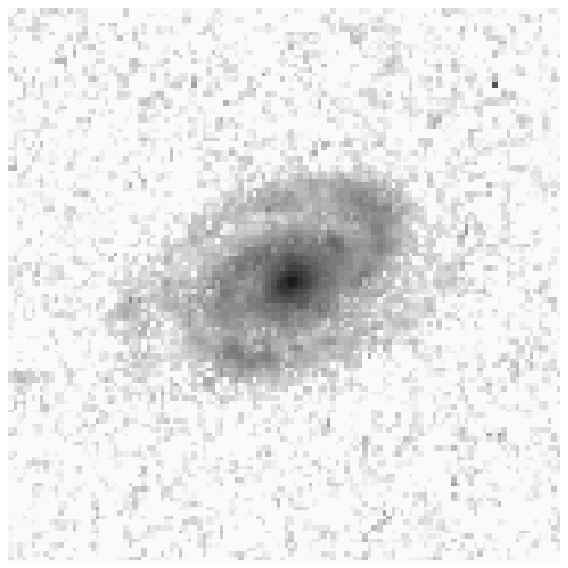,width=5.8cm,clip=}
   \hspace{0.0cm}
    \psfig{figure=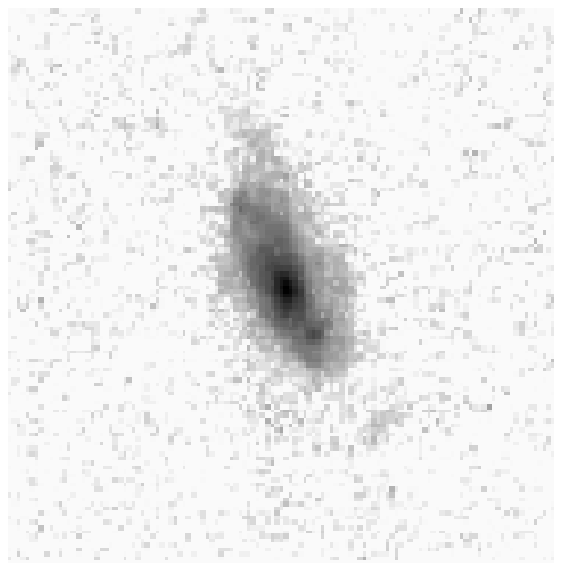,width=5.8cm,clip=}
   \hspace{0.0cm}
    \psfig{figure=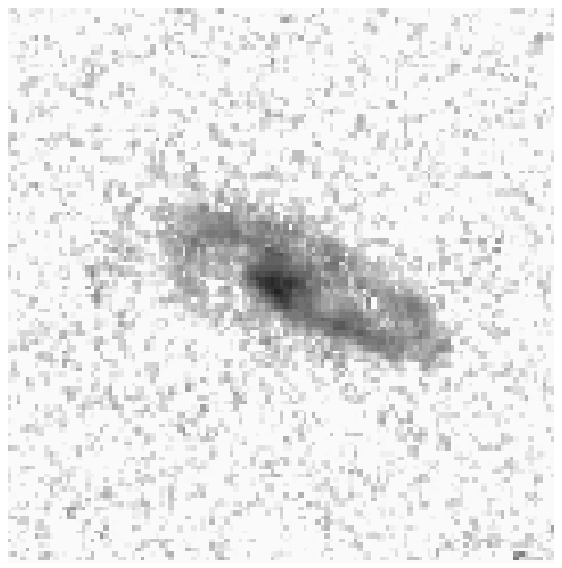,width=5.8cm,clip=}}
   \vspace{0.1cm}              
   \hbox{\hspace{0cm}              
    \psfig{figure=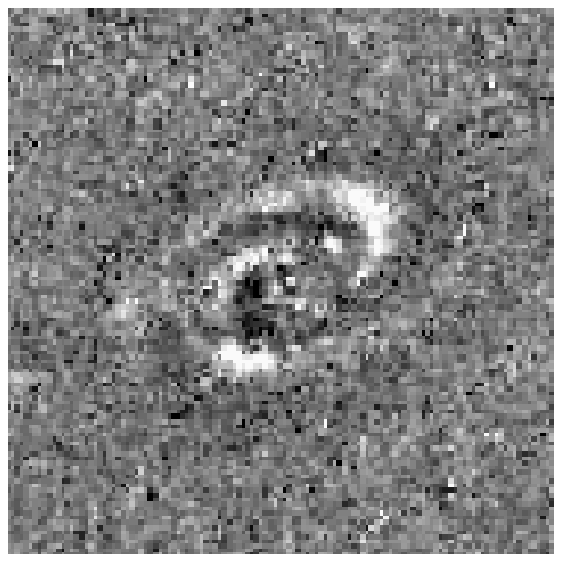,width=5.8cm,clip=}
   \hspace{0.0cm}
    \psfig{figure=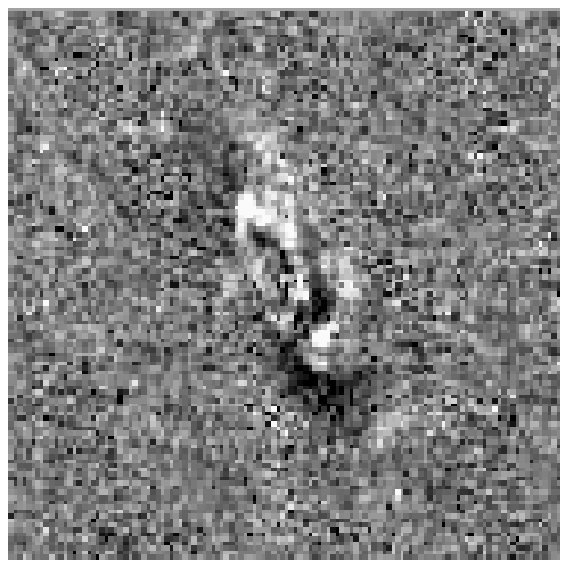,width=5.8cm,clip=}
   \hspace{0.0cm}
    \psfig{figure=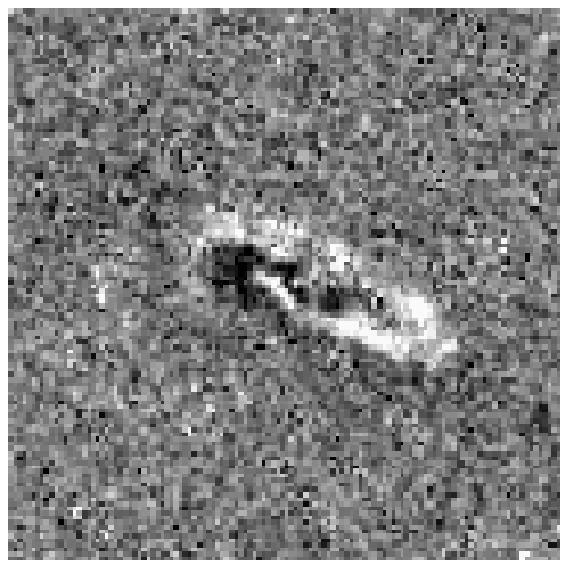,width=5.8cm,clip=}}}
  \caption[]{{\bf Upper panel:} HST images of FDF\,2484 ($z$ = 0.65), 
FDF\,2174 ($z$ = 0.68), and FDF\,4922 ($z$ = 0.97). North is up and east 
is left. The scale is 0.05 arcsec/pix, the image size is $6" \times 6"$.\\
{\bf Lower panel:} Corresponding residual images after subtraction
of 2-dim models for disk and bulge. The remaining structures show the
dominant bisymmetric arms.  White, grey, and black colors corresponds to
positive, zero, and negative residuals, respectively}. 

  \label{hstimag}
   \end{figure*}

In the present paper we investigate spiral galaxies identified in the FORS Deep
Field (hereafter referred to as FDF, Appenzeller et al. 2000). 
The photometric survey is described in detail by Heidt et al.~(2003). 
Ziegler et al.~(2002) and B\"ohm et al.~(2004) observed the kinematics of  
a large sample of FDF-galaxies in the redshift range $z = 0.1 - 1$. 
Three of the galaxies at $z > 0.5$ turn out to be suitable for a quantitative 
analysis of their morphologies and kinematics and allow to extend our previous
studies out  to redshifts $z \approx 1$. The data are described in section 2. 
In sections 3 and 4 we construct dynamical models for the galaxies. As 
discussed in the final section the dynamical modelling seems to indicate 
maximum disks also in field galaxies in that redshift range.

\begin{table}
\caption[]{Redshifts, absolute rest frame magnitudes, \\ seeing (FWHM), 
maximum rotation velocities, 
\\ and scale conversion factors of the observed galaxies}
         \label{see}
\[
\begin{tabular}{cccccc}
\hline
 \noalign{\smallskip}
  FDF & redshift & $M_{\rm B}$ & seeing & $V_{\rm max}$ & scale \\ 
       & & mag & arcsec & km/s & kpc/arcsec \\
 \noalign{\smallskip}
 \hline
 \noalign{\smallskip}
  2484 & 0.6535 & -21.07 &0.80 & 169 & 6.14 \\
  2174 & 0.6798 & -21.00 &0.51 & 175 & 6.26 \\
  4922 & 0.9731 & -21.78 &0.43 & 156 & 7.20 \\
 \noalign{\smallskip}
 \hline
 \end{tabular}
 \]
\end{table}

\section{Observations}

\begin{figure*}
  \vbox{
   \hbox{\hspace{0cm}              
    \psfig{figure=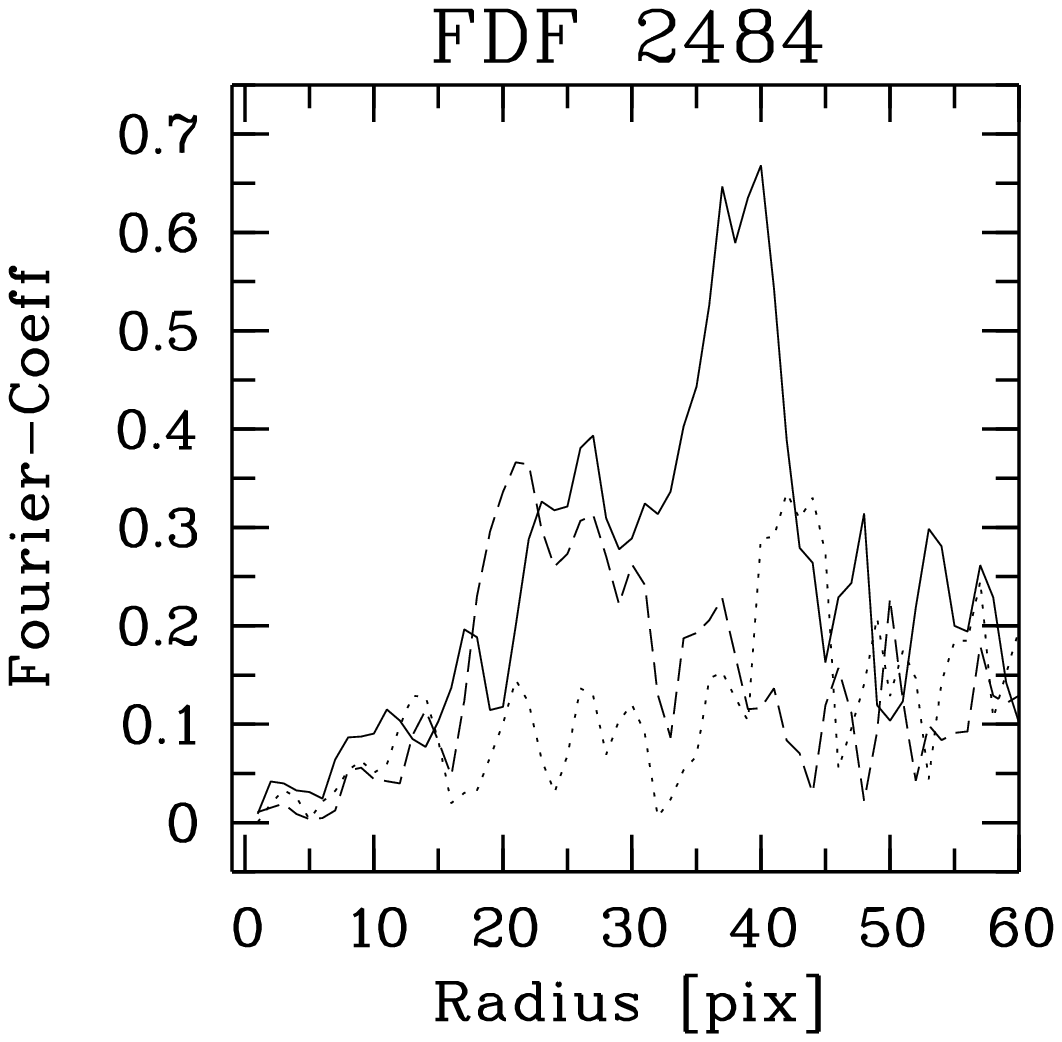,width=6.0cm,clip=}
   \hspace{0.0cm}
    \psfig{figure=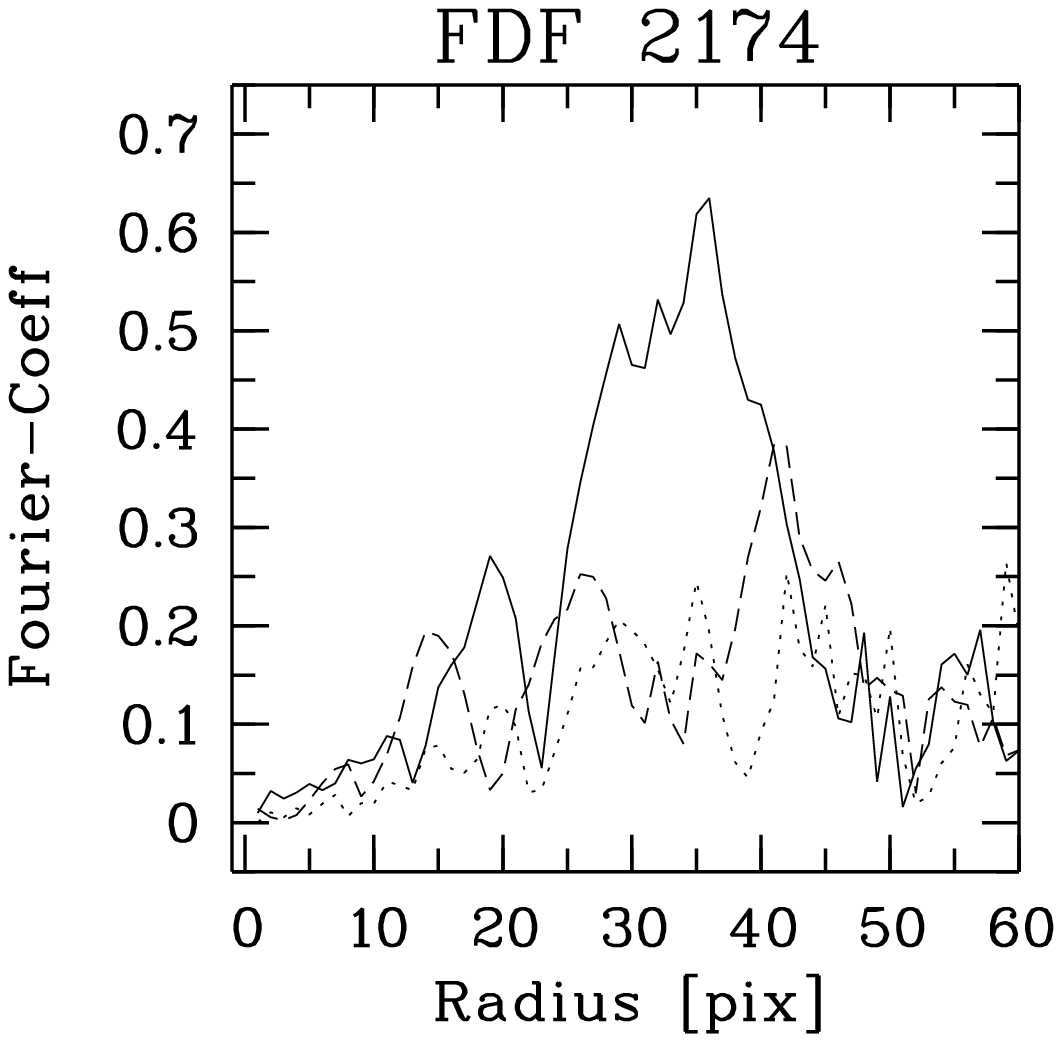,width=6.0cm,clip=}
   \hspace{0.0cm}
    \psfig{figure=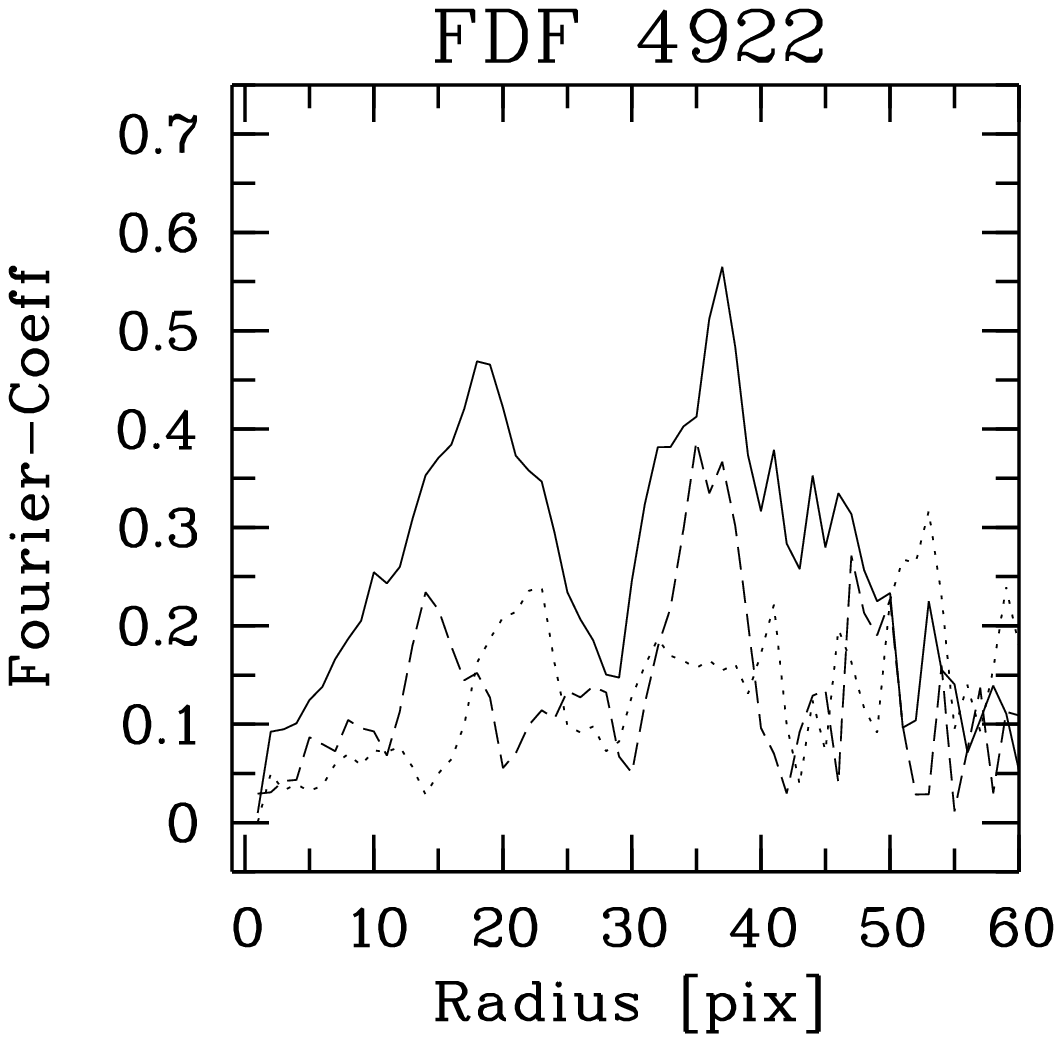,width=6.0cm,clip=}}}
  \caption[]{
Fourier coefficients (in arbitrary units) $m = 2$ (solid line), 
$m = 3$ (dashed line), and $m = 4$ (dotted line) of the spiral arm patterns in  
FDF\,2484, FDF\,2174, and FDF\,4922, respectively. The coefficient $m = 2$ is
dominant in all cases.}
  \label{fourcomp}
   \end{figure*}

B\"ohm et al.~(2004) have measured spatially resolved rotation 
curves of 77 galaxies in the FDF using the FORS spectrograph at VLT.
36 of these rotation curves were classified as high quality in terms of
large radial extent and small asymmetries. Moreover, Ziegler and collaborators 
(B\"ohm et al., in preparation) have obtained HST images of the FORS Deep Field
with the ACS wide field camera. 
The $\sim$ 6 $\times$ 6 arcmin field--of--view of the
FDF was observed within 4 visits. Each visit was
splitted into two exposures of 1180\,s integration time using the F814W filter.
We have examined the ACS image of each high quality object 
in the redshift range $z = 0.5 - 1$
and found three galaxies showing discernible spiral structure:
FDF 2484 at redshift $z \approx 0.65$, 
FDF 2174 at $z \approx 0.68$, and FDF 4922 at $z \approx 0.97$, respectively. 
The HST images of the galaxies are reproduced in the upper panels of 
Fig.~\ref{hstimag}. As can be seen from the residual images in the lower panels
all galaxies show well developed two-armed spirals. The $m = 2$ arm mode was
confirmed by a Fourier analysis of the arm pattern as described in the
next section (Fig.~\ref{fourcomp}).

%

\begin{table}
 \caption[]{Photometric parameters from the bulge-disk decompositions.
  $\Sigma_{\rm b0}$ and $\Sigma_{\rm d0}$ are the projected central 
  flux-densities
  of bulge and disk, measured in analog-digital units (adu) per pixel.
  $R_{\rm b}$ and $R_{\rm d}$ are the characteristic scale lengths.
  $i$ is the inclination of the disk and $\theta$ is the angle 
  between spectrograph slit and apparent major axis of the galaxy.} 

   \label{pho}
   \[
   \begin{tabular}{rrrcccr}
            \hline
            \noalign{\smallskip}
            FDF & $\Sigma_{\rm b0}$ & $\Sigma_{\rm d0}$ &
            $R_{\rm b}$ & $R_{\rm d}$ & $i$ & $\theta$\\
            & adu/pix & adu/pix & arcsec & arcsec & deg & deg \\
            \noalign{\smallskip}
            \hline
            \noalign{\smallskip}
             2484 & 157 & 133 & 0.13 & 0.65 & 54 & -12\\
             2174 & 182 & 251 & 0.12 & 0.50 & 67 & -4\\
             4922 & 64 & 62 & 0.22 & 0.82 & 71 & 14 \\
            \noalign{\smallskip}
            \hline
         \end{tabular}
      \]
   \end{table}
  
The optical rotation curves are shown as data points in
Figs.~\ref{rc2484} to \ref{rc4922} in section 4. 
The seeing during the spectroscopic observations is summarized in Table 1. 
For the determination of the misalignment between the galaxies'
apparent major axes and the slit directions, we took into account
a slight rotation of the HST/ACS frames with respect to the FORS spectra.
This tilt has been derived by coordinate transformations between the two
reference systems and amounts to approximately $-4$\,deg. The last column in 
Table 1 gives scale conversion factors from arcsec 
to kpc. These have been determined in the same way as by Vogt et al.~(1996) 
adopting a cosmological model with $q_0 = 0$, which 
at the redshifts discussed here is similar to the 
concordance model with $\Omega_{\rm m} = 0.3$ and $\Omega_\Lambda = 0.7$. 
We assumed a Hubble constant of $H_0 = 75$ km s$^{-1}$ Mpc$^{-1}$.

As can be seen from Figs.~3 to 5 the rotation curves have been measured
with good accuracy. The methods to extract the rotation curves from the spectra
and to estimate the errors are described in detail in B\"ohm et al.~(2004). The
rotation curves are very regular and quite symmetric. Taken together
with the unperturbed optical appearances of the galaxies this should allow in
our view a reliable comparison with model rotation curves in order to draw
inferences about the dark matter content.

\section{Bulge--disk decomposition}

The surface-brightness distributions of the galaxies were modelled in the 
same way as in Fuchs, M\"ollenhoff \& Heidt (1998). 
A thin disk was assumed which has elliptical isophotes when seen under
an inclination angle $i$. The radial brightness profiles were fitted by an
exponential function 
\begin{equation}
\Sigma_{\rm d}(R) = \Sigma_{\rm d0} \exp{(-R/R_{\rm d})}
\end{equation}
with a scale length $R_{\rm d}$. For reasons of the easier handling of the
dynamical models we adopt instead of a de Vaucouleurs law for the surface 
brightness profiles of the bulges softened power laws
\begin{equation}
\Sigma_{\rm b}(R) = \Sigma_{\rm b0} \left( 1 + \frac{R^2}{R_{\rm b}^2} 
\right)^{-{2.5}/{2}}\,,
\end{equation}
where $R_{\rm b}$ denotes the core radius. The corresponding 
three--dimensional distribution has the same form as equation (2), but with an
exponent lowered to -3.5/2. The central density of the three-dimensional
distribution $\rho_{\rm b0}$ is given by $\rho_{\rm b0} = 0.572\,
\Sigma_{\rm d0}/R_{\rm b}$. The models (1) and (2) were 
fitted simultaneously to the data using the
two--dimensional fitting code of M\"ollenhoff (2004). The parameters  
$\Sigma_{\rm b0}, \Sigma_{\rm d0}, R_{\rm b}, R_{\rm d},i$, and the position 
angle of the major axis of the disk are free fit-parameters. 
The photometric parameters of the best fits are given in Table 2.
The angle $\theta$ in Table 2 is the position angle of the slit relative to the
major axis of the inclined galaxy. 

The fit to the surface--brightness distribution describes the axisymmetric
structure of disk and bulge. Therefore the spiral arms appear clearly
in the residual image = galaxy -- model (Fig.~\ref{hstimag}, lower panels).
The residual images were transformed
to face--on view and the spiral arm pattern was Fourier-analyzed.
Fig.~\ref{fourcomp} shows the radial variation of the coefficients
$m$=2, 3, 4 in the galaxies. The two--armed pattern $m$=2 is dominant
in all cases. The image of FDF\,2174 shows that the spiral
extends radially out to about 1.3 arcsecs. This corresponds in the diagram of 
the $m$=2 Fourier coefficients in Fig.~\ref{fourcomp} to the indentation at 
a radius of about 25 pixel lengths. 
The very large amplitude of the $m$=2 Fourier coefficients beyond that
radius is caused by two outer patches lying nearly diagonally towards northeast
and southwest, which are, however in our view, 
not related to the spiral structure (cf. Fig.~\ref{hstimag}).  
But as evidenced by the rotation curve, these outer regions do belong
to the optical disk of the galaxy.

\section{Decomposition of the rotation curves}

We adopt model rotation curves of the form
\begin{equation}
\upsilon_{\rm c}^2(R) = \upsilon_{\rm c,b}^2(R) + \upsilon_{\rm c,d}^2(R) 
+ \upsilon_{\rm c,h}^2(R)\,,
\end{equation}
where $\upsilon_{\rm c,b}$, $\upsilon_{\rm c,d}$, and $\upsilon_{\rm c,h}$ 
denote the contributions by the bulge, disk, and dark halo, respectively. The 
bulge contribution is given by
\begin{equation}
\upsilon_{\rm c,b}^2(R) = \frac{4 \pi G \rho_{\rm b0}}{R} \int_0^R dr r^2
\left( 1 + \frac{r^2}{R_{\rm b}^2} \right)^{-{3.5}/{2}} \,,
\end{equation}
where $G$ denotes the constant of gravitation. The rotation curve of an
infinitesimally thin exponential disk is given by Freeman's (1970) formula,
\begin{equation}
\upsilon_{\rm c,d}^2(R) = 4 \pi G \Sigma_{\rm d0} R_{\rm d} x^2 (\,I_0(x)
 K_0(x) - I_1(x) K_1(x)\, ) \,,
\end{equation}
where $x$ is an abbreviation for $x = R/2R_{\rm d}$ and $I$ and $K$ denote
Bessel functions. In equations (4) and (5) we use the 
scale lengths determined in the previous section and the ratios of
$\rho_{\rm b0}$ /$\Sigma_{\rm d0}$ found by the quantitative modelling of
the surface brightness photometry. Only the mass--to--light ratio is a free 
parameter which is determined as explained below. The dark halos are modelled
as quasi--isothermal spheres
\begin{equation}
\rho_{\rm h}(R) = \rho_{\rm h0} \left( 1 + \frac{r^2}{R_{\rm h}^2} 
\right)^{-1} \,,
\end{equation}
which leads to a contribution to the rotation curve of the form
\begin{equation}
\upsilon_{\rm c,d}^2(R) = 4 \pi G \rho_{\rm h0} R_{\rm h}^2 \left(
 1 - \frac{R_{\rm h}}{R} \arctan{\left( \frac{R}{R_{\rm h}} \right)} \right) \,.
\end{equation}
Such density profiles of dark halos with homogeneous cores fit usually the
rotation curves of galaxies better (de Blok, McGaugh \& Rubin 2001, de Blok 
\& Bosma 2002, Gentile et al.~2004) than centrally cusped profiles as
proposed, for instance, by Navarro, Frenk \& White (1997).
The velocity fields described by the rotation curve model (3) are projected
onto the sky adopting the inclination angles determined in the previous
section. Next, the line--of--sight components of the rotation velocities are
calculated and weighted by the surface brightness.
The radial velocity fields were then convolved with
two--dimensional Gaussians in order to model the blurring of the velocity fields
by seeing. Finally the velocity field models were masked according to a slit
width of 1 arcsec and the position angles of the slit given in Table 2 (cf.
also Fig.~4 of B\"ohm et al. 2004). The synthetic rotation curves determined
this way can be directly compared with the observed rotation curves.

Since the decomposition of the rotation curves is completely degenerate, we set
constraints on the decomposition using arguments of the density wave theory of
galactic spiral arms. The density wave theory predicts that spiral density 
waves grow preferentially with azimuthal wavelengths (Toomre 1981, Fuchs 2001)
\begin{equation}
\lambda = X(A) \lambda_{\rm crit} = X(A) \frac{4 \pi^2 G \Sigma_{\rm d}}
{\kappa^2} \,,
\end{equation}
where $\kappa$ denotes the epicyclic frequency of the stellar orbits,
$\kappa = \sqrt{2} (\upsilon_{\rm c}/R) \sqrt{1 +
 {d\,ln\,\upsilon_{\rm c}/}{d\,ln\,R}}$. 
The coefficient $X$ depends on the slope of the rotation curve
measured by Oort's constant $A$ (Fuchs 2001). For a flat rotation curve its 
value is $X = 2$. The expected number of spiral arms $m$ is obviously given by
how often the circumferential wavelength $\lambda$ fits onto the annulus,
\begin{equation}
m = \frac{2 \pi R}{\lambda} \,.
\end{equation}
As was pointed out first by Athanassoula (1988) and Athanassoula et al. (1987)
equations (9) and (8) can be used to determine the surface density of galactic
disks once the rotation curve $\upsilon_{\rm c}(R)$ and thus the epicyclic 
frequency $\kappa$ are known. In this way the morphological appearance of 
spiral galaxies constrains the decomposition of the rotation curve into its 
various contributions.

\begin{table}
  \caption[]{Dynamical parameters from the decomposition of the 
  rotation curves. $\rho_{\rm b0}$ is the  central spatial density of the bulge
  (in ${\mathcal{M}_\odot}/{pc^3}$). $\Sigma_{\rm d}$ is the central
  surface density of the disk (${\mathcal{M}_\odot}/{pc^2}$), and 
  $\rho_{\rm h0}$ is the central spatial density of the halo. The next 3 columns
  give the mass of bulge, disk, and halo, within radii 12, 10, and 15 kpc for 
  FDF\,2484, 2174, and 4922, resp. The last column is the
  resulting rest frame B mass-to-light ratio calculated using total disk mass
  estimates.}  
         \label{dyn}
      \[
         \begin{tabular}{rrrrrrrr}
            \hline
            \noalign{\smallskip}
            FDF & $\rho_{\rm b0}$ & $\Sigma_{\rm d}$ &
            $\rho_{\rm h0}$ & $\mathcal{M}_{\rm b}$ & 
            $\mathcal{M}_{\rm d}$ & $\mathcal{M}_{\rm h}$ & 
            $\frac{\mathcal{M}}{\mathcal{L}}$\\
            \noalign{\smallskip}
            &  $\frac{\mathcal{M}_\odot}{pc^3}$
            &  $\frac{\mathcal{M}_\odot}{pc^2}$
            &  $\frac{\mathcal{M}_\odot}{pc^3}$
            & \multicolumn{3}{c}{ $ 10^9 \mathcal{M}_\odot $}
            &  $\frac{\mathcal{M}_\odot}{\mathcal{L}_{{\rm B}\odot}} $\\
            \noalign{\smallskip}
            \hline
            \noalign{\smallskip}
             2484 & 0.41 & 480 & - & 4.6 & 39 & - & 1.5\\
                  & 0.38 & 450 & 0.0016 & 4.6 & 36 & 6.5 & 1.4\\
            \noalign{\smallskip}
             2174 & 0.38 & 700 & - & 3.7 & 36 & - & 1.4\\
                  & 0.33 & 600 & 0.002 & 3.7 & 31 & 5.4 & 1.2\\
           \noalign{\smallskip}  
             4922 & 0.13 & 350 & - & 11 & 55 & - & 1.3 \\
                  & 0.12 & 310 & 0.0013 & 11 & 49 & 8.4 & 1.2 \\
            \noalign{\smallskip}
            \hline
         \end{tabular}
      \]
   \end{table}

We have tested the prediction of equation (9) on the
numerical simulation of the dynamical evolution of a galactic disk by Sparke \&
Sellwood (1987). The mass of the disk and the rotation curve are known from the
setup of the simulation and we find that equation (9) predicts indeed a
two--armed spiral as seen in the first snapshots of the simulation before it
develops a bar.

\section{Results and discussion}

\begin{figure}
  \centering
  \includegraphics[width=8cm]{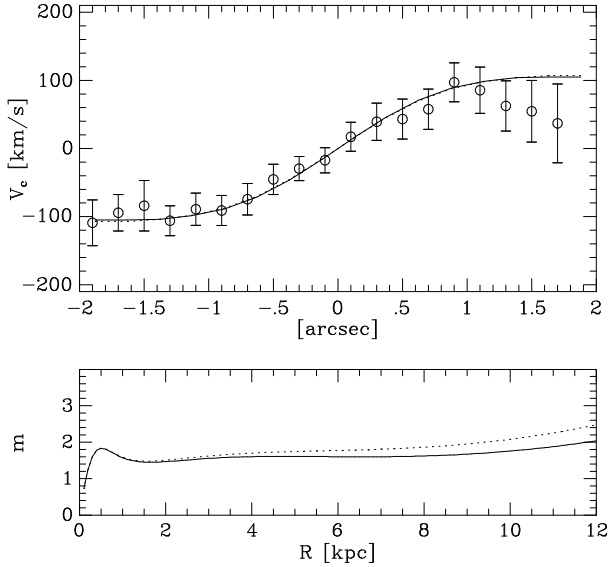}
  \caption[]{{\bf Upper panel:} Model rotation curves of FDF\,2484 fitted to the
radial velocity data of B\"ohm et al.~(2004). The solid line is the
model without a dark halo, the dotted line shows the 
corresponding rotation curve when a dark halo component was included.
The rotation curves are practically identical.\\  
{\bf Lower panel:} Expected number of spiral arms in the radial range where 
spiral structure can be seen. The dotted curve is again for the model
with a dark halo included.}
  \label{rc2484}
  \end{figure}

\begin{figure}
   \centering
   \includegraphics[width=8cm]{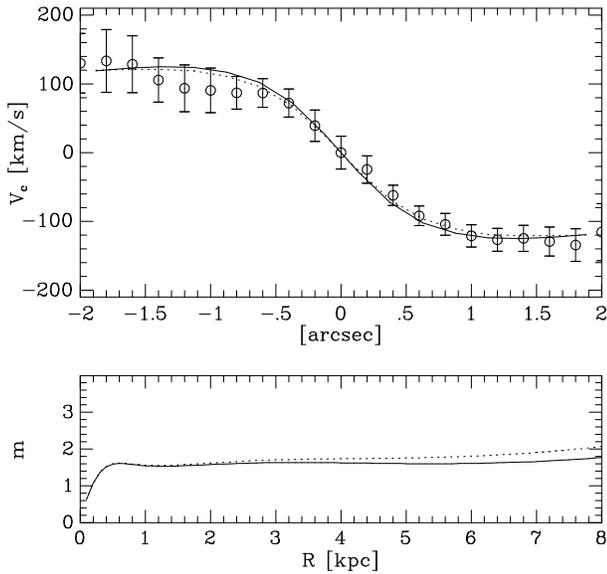}
   \caption[]{Same as Fig.~\ref{rc2484}, but for FDF\,2174.}
   \label{rc2174}
   \end{figure}

 \begin{figure}
   \centering
   \includegraphics[width=8cm]{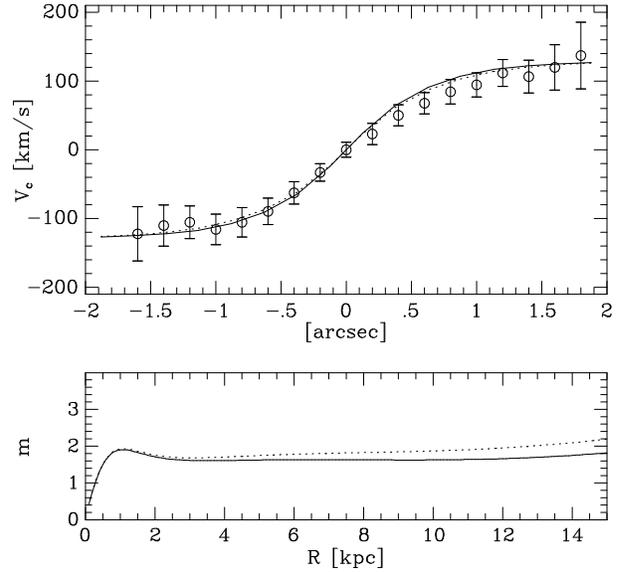}
   \caption[]{Same as Fig.~\ref{rc2484}, but for FDF\,4922.}
   \label{rc4922}
   \end{figure} 

In the upper panels of Figs.~\ref{rc2484} to \ref{rc4922} we present 
model rotation curves fitted to the observed
rotation curves of the three galaxies assuming no dark halos. As can be seen
from the figures the models fit well to the data. The corresponding dynamical
parameters are listed in Table 3. The expected number of spiral arms estimated
using equation (9) are shown as solid lines in the lower panels of 
Figs.~\ref{rc2484} to \ref{rc4922}. In each case
the prediction is a two--armed spiral, exactly as observed. 
This hardly allows massive inner dark halos. Nevertheless, we also performed
fits with dark halos.

Unfortunately, the rotation curves span radially only between two and four 
disk scale--lengths. The core 
radii of the dark halos are not well constrained in such cases (van der Kruit 
1995). We have therefore adopted rather arbitrarily for each galaxy a core 
radius of 10 kpc. The central halo densities were chosen so that the expected 
number of spiral arms is still in accordance with the observed morphologies of 
the galaxies, i.e.~two--armed spirals. The resulting fits are shown 
as dotted lines in Figs.~\ref{rc2484} to \ref{rc4922}. As can be seen from the
figures they are of the same quality as in the case of the bulge--disk models 
without dark matter. The dark halo contributions could be
raised and simultaneously the disk contributions lowered further, and would lead
to equally good fits to the rotation curves. However, the expected number of 
spiral arms would then increase according to equation (9) as $m \propto 
\Sigma_{\rm d}^{-1}$ and would fit no longer to the observed morphologies of the
galaxies. 
Thus, applying the density wave theory constraint to the decomposition of the 
rotation curves the disks of the three galaxies in the $z = 0.7$ to 1 redshift
range we find that the galaxies seem to have maximum disks similar to the 
nearby galaxies. This does not mean that we argue against the presence of dark 
matter in the galaxies. The observed rotation curve of FDF\,2174, for example, 
spans already about four radial scale lengths and shows no indications of 
falling at the outer radii. This would be the case if there were only a single 
disk. Instead the flat shape of the rotation curve implies according to our 
model a ratio of $\mathcal{M}_{\rm h}/(\mathcal{M}_{\rm b} + 
\mathcal{M}_{\rm d})$ = 0.24 at galactocentric radius $R$ = 12.5 kpc.

Using the absolute rest frame luminosities of the FDF galaxies derived by 
B\"ohm et al.~(2004), transformed to the value of $H_0$ adopted here, we have
determined the
mass--to--light ratios shown in Table 3. The dynamically determined
mass--to--light ratios are in good agreement with stellar 
mass--to--light ratio estimates based on population synthesis models for 
galaxies in this this redshift range (Dickinson et al.~2003, Drory et al.~2004).
   
We have demonstrated that there exist galaxies out at redshifts of $z \approx 1$
which are dominated by baryons in their inner parts. Obviously this must 
reflect on theories of the cosmogony of the galaxies. It would be interesting 
if the present paradigm of $\Lambda$ cold dark mater cosmology can account for
such galaxies. For instance, the recent state--of--the--art simulation of 
galaxy formation by Abadi et al.~(2003) lead to a galaxy model dominated by dark
matter through all phases of its evolution, which is at variance to our 
findings. Admittedly the selection criterion of visible spiral structure might
introduce a bias in the galaxy sample which we have analyzed. Since the 
galaxies have organized themselves quite early by self gravity, they have 
probably not experienced a major merger event and could settle into 
equilibrium.   

\begin{acknowledgements}
      This work was supported by the 
      \emph{DFG Sonderforschungsbereich} 439,
      the \emph{Volkswagen Foundation} (I/76\,520) and the \emph{
      Deutsches Zentrum f\"ur Luft- und Raumfahrt} (50\,OR\,0301).
      We are grateful for the comments by the anonymous referee which 
      clarified the presentation of the paper.
\end{acknowledgements}


\end{document}